# Cavity-induced backscattering in a two-dimensional photonic topological system

Yuhao Kang[1,2], Azriel Z. Genack[1,2*]

[1]Department of Physics, Queens College of the City University of New York,

Flushing, NY 11367, USA

[2]Graduate Center of the City University of New York, New York, NY 10016,

USA

Email: [*]genack@qc.edu

Abstract:

The discovery of robust transport via topological states in electronic, photonic and phononic materials has deepened our understanding of wave propagation in condensed matter with prospects for critical applications of engineered metamaterials in communications, sensing, and controlling the environment. Topological protection of transmission has been demonstrated in the face of bent paths and on-site randomness in the structure. Here we measure the propagation of microwave radiation in a topological medium possessing time reversal symmetry with a cavity adjacent to the edge channel. A coupled-resonance model analysis shows that the cavity is not a spin-conserving defect and gives rise to negative time delay in transmission.

**1. Introduction**

Recent progress in the study of photonic topological structures has opened new possibilities for robust wave transport from microwave to optical frequencies [1,2].Topologically protected edge states, whose progress is undeterred by point defects or bent paths propagate along the boundary between topological and trivial domains with different topological invariants [3].

Several schemes have been proposed to emulate the quantum spin Hall (QSH) effect and test the robustness of edge modes. Hafezi et al. [4] propose a network of coupled resonator optical waveguides in which the pseudo-spin is emulated by clockwise and counterclockwise circulation in the ring. They showed in numerical simulations that transmission through the edge mode remains near unity in the face of random fluctuations in resonator frequencies as the system size increases. QSH-like topological insulators (TIs) were also realized in bi-anisotropic metamaterials [5,6]. The measured time delay of edge modes averaged over disorder is consistent with ballistic transport time [7].

In addition to perfect transmission, the ability to control the time delay in transmission and in paths within the TI could be important in many applications. The time delay can be modified by either creating localized modes inside the bulk region by introducing disorder or by changing the length of the boundary line between domains. The impact of defects is also of interest because no system is perfect and there may be defects in a topological structure that support discrete modes in the vicinity of the edge. The random disorder that will be considered is not strong enough that the band gap in the nontrivial system begins to close. Several different types of disordered have been considered, including highly-correlated disordered crystals [8] and random fluctuations in onsite energy [9]. An immediate question is whether unidirectional propagation of edge mode persists in a large-scale of disorder in which localized mode are created within the bulk bandgap. This is of particular interest because it is possible to engineer the type of disorder and its proximity to the edge in order to slow down an edge mode by coupling to such states.

The magnitude and phase of the field transmitted through a continuum channel coupled to a discrete mode is a Fano resonance. This approach to scattering was first proposed to explain the asymmetric spectrum of the electron-molecule scattering cross section [10]. The interference between the discrete autoionized state and the freely propagating continuum gives rise to a characteristic asymmetric shape in the scattering spectrum. This phenomenon can be extended to classical interference between a discrete state and a slowly changing background. The sharp Fano resonance in transmission has many applications in the photonic crystal waveguide structures, such as an optical switch from complete transmission to complete reflection [11]. Here we observe a Fano resonance in a non-trivial photonic crystal. A cavity introduced inside the bulk region is found to support several discrete modes within the band gap. We will analyze the observed spectrum by generalizing the standard Fano formula to the case of multiple-modes using coupled-mode theory. Understanding the coupling between the edge mode and modes of a cavity is a first step in tailoring transport in TIs via cavity modes.

We study a TI with TR symmetry. This system does not require real or auxiliary magnetic fields to realize topological protection in bent paths and weak onsite distributed disorder. The sample has a topological domain with the triangular lattice shown in the upper half of Fig. 1a adjoining a trivial domain with the triangular lattice shown in the lower half of the figure. The topological lattice is composed of rods with a concentric collar. The position of the collar can be changed by pushing the rod, which protrudes through the holes drilled in the two plates. The waveguide supports both transverse-electric-like (TE) and transverse-magnetic-like (TM) modes. When the collar is midway between the plates, the TE and TM modes are degenerate at the Dirac point. Pseudo-spin-up and spin-down states are the bonding and antibonding combinations of these modes. TE and TM modes are coupled when the collar is displaced from a midpoint between the plates, effectively emulating spin orbit coupling in electronic systems and leading to topological order. This system can be described by the Kane-Mele model $H = v(\delta k_x \tau_z s_0 \sigma_x + \delta k_y \tau_0 s_0 \sigma_y) + m\tau_z s_z \sigma_z$, where the Pauli matrices $\hat{\tau}$, $\hat{s}$ and $\sigma$ act on the subspace of valley, spin, and double states respectively. $v$ is the group velocity, and $m$ is the mass term due to bi-anisotropy [5]. This Hamiltonian is a good approximation near the Dirac point. Pushing the rod to the opposite plate changes the sign of $m$. The electromagnetic wave is confined to the boundary between the 2D lattice of rods and collars and a 2D array of triangular prisms with a gap between them, which form a trivial insulator with bandgap coinciding with that of the adjacent TI [12,13]. Two pseudo-spin-polarized edge modes propagate in opposite directions along the interface to form a Kramers pair. To create a cavity, seven rod-collar units are pushed up so that the collars are in contact with the upper plate.

In an electronic TI system with TR symmetry, the Kramers theorem ensures the decoupling of the single pair of helical edge states [14]. In a photonic system, previous theoretical, numerical, and experimental studies show spin flipping was inhibited and reflection was suppressed. However, the only disorder discussed so far has been introduced via point defects or bent paths. The effect of a large cavity near the edge channel on the reflection rate has not been studied previously in a real structure.

In this work, we carry out microwave measurements of wave transport through a QSH TI in which a single defect cavity is introduced adjacent to the edge. We extract the reflection through a fit of an analytical model of the Fano resonance formed by the cavity and continuum edge mode to the measurements of field spectra inside the cavity.

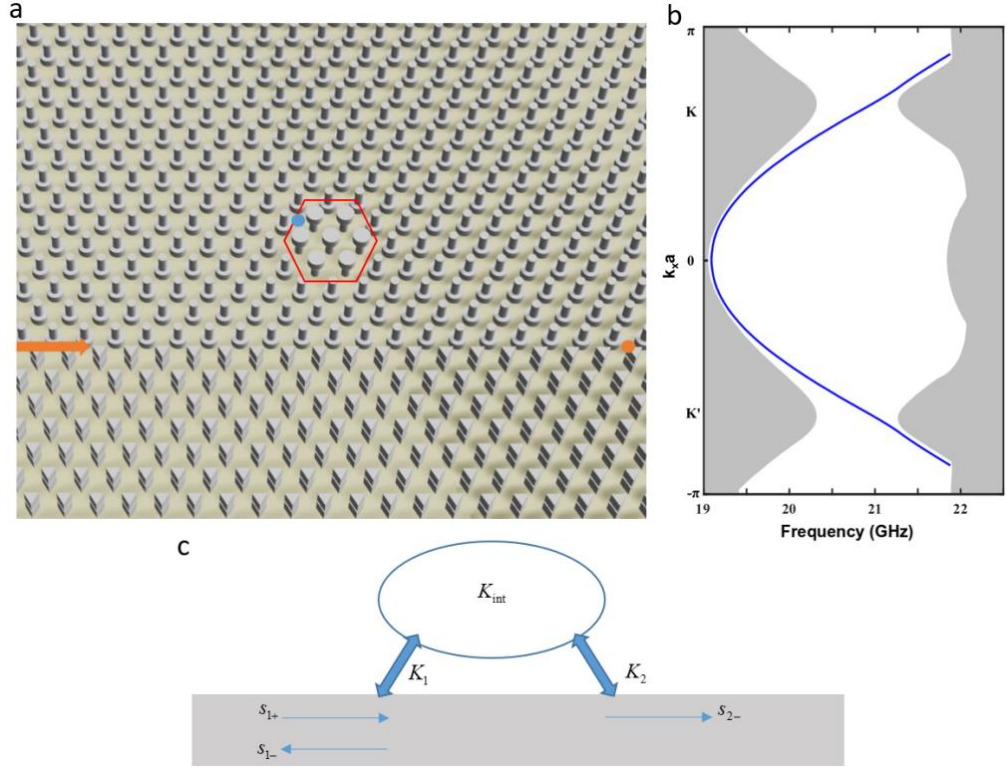

Fig.1. Schematic diagram of the structure. (a) Structure without the upper plate. The red arrow indicates the direction of flow of the incident wave. The blue and orange points are the positions of two detectors located within the cavity and at the output of the sample, respectively. The sample parameters are given in [7]. They are: Lattice constant: 1.0890 cm, height of rod: 1.0890 cm, height of collar; 0.3580 cm, diameter of collar: 0.6215 cm, diameter of rod: 0.3175 cm, height of triangular prism: 0.5040 cm, gap between triangular prisms: 0.0810 cm, and side length of triangular prism: 0.5020 cm. Within the red border, rods are pushed up so that their attached collars contact the upper plate. The real structure contains 15 layers of triangular prisms and 25 layers of rods. (b) Band structure of edge mode. The bulk bandgap extends from 20.3 to 21.3GHz. (c) Schematic of coupled-resonator model. $K_1$/$K_2$ is the coupling coefficient between the cavity mode and the left/right side of the edge channel. $K_{int}$ denotes the internal loss of the cavity.

## 2. Coupled-mode theory

We treat multiple modes and internal loss in the cavity using coupled mode theory [15,16]. Complex field of the $n$th mode $a_n$ with central frequency $\omega_n$, $a_n = ae^{-i\omega t}$ evolve as

$$\frac{da_n}{dt} = (-i\omega_n - \Gamma_{n1} - \Gamma_{n2} - \Gamma_{n\text{int}})a_n + K^T |s_+\rangle$$
$$|s_-\rangle = C|s_+\rangle + Da \quad (1)$$

Here $C = \begin{bmatrix} r & t' \\ t & r' \end{bmatrix}$ is the scattering matrix directly through the continuum state, $a = (a_1, a_2 \cdots a_n)^T$, $\Gamma_{n1}, \Gamma_{n2}, \Gamma_{n\text{int}}$ denote the decay rates of the $n$th cavity mode due to coupling to the left and right and internal loss, respectively. $K$ and $D$ are the coupling matrices between discrete modes and the incoming and outgoing ports, respectively.

$K = \begin{bmatrix} K_{11} & K_{21} & \cdots & K_{n1} \\ K_{12} & K_{22} & \cdots & K_{n2} \end{bmatrix}$, $K_{n1/2}$ is the strength of coupling between the $n$th cavity and port 1/2. $|K_{n1/2}|^2 = 2\Gamma_{n1/2}$ [16]. Due to electromagnetic reciprocity, $D = K$ [17].

$|s_\pm\rangle = (s_{1\pm}, s_{2\pm})^T$, with $s_{2+} = 0$. From Eq. (1), we find $a_n = \dfrac{K_{n1} s_{1+}}{i(\omega_n - \omega) + \Gamma_{n\text{tot}}}$,

$s_{2-} = t s_{1+} + \sum_n K_{n2} a_n$. The transmission coefficient can then be expressed as

$$\tau = \frac{s_{2-}}{s_{1+}} = t + \sum_n \frac{K_{n1} K_{n2}}{i(\omega_n - \omega) + \Gamma_{n\text{tot}}}, \quad (2)$$

where $\Gamma_{n\text{tot}} = \Gamma_{n1} + \Gamma_{n2} + \Gamma_{n\text{int}}$ is the total linewidth of $n$th mode.

## 3. Experimental results and spectral analysis

To determine the linewidth and central frequency of the cavity modes in this system, we measure the spectrum of the field inside the cavity. The probe is placed at the upper-left corner of the cavity, which is indicated by the blue point of Fig. 1a. The position of probe relative to the cavity is fixed in the following measurements. The results for separation between the cavity and boundary of 1, 3, 5, and 7 layers are shown in Fig. 2. The red lines in the upper panels of Figs. 2a-d are the measured signals, which are the superposition of the $a_n$ obtained from Eq. (1)

$$E(\omega) = \sum_n \frac{V_n}{\omega - \omega_n + i\Gamma_{n\text{tot}}}. \quad (3)$$

Here $V_n$ describes the coupling between the cavity and edge modes, which depends on the detector position and field distributions of the cavity modes.

The parameters $\omega_n$ and $\Gamma_{n\text{tot}}$ are retrieved from the modal decomposition of the spectrum using the method of harmonic inversion [18]. The dotted blue curves in the upper panels of Figs. 2a-d show the fit to the dotted curves of the measured spectrum. The lower panels in these figures show the contribution of each mode between 19.5 and 22.5 GHz. The coupling strength decreases when the cavity is moved further from the edge. Table 1 lists the central frequencies and linewidths extracted from the measurements in the range from 20.3 and 21.3 GHz. The first column gives the eigenfrequencies of modes solved using COMSOL mode solver.

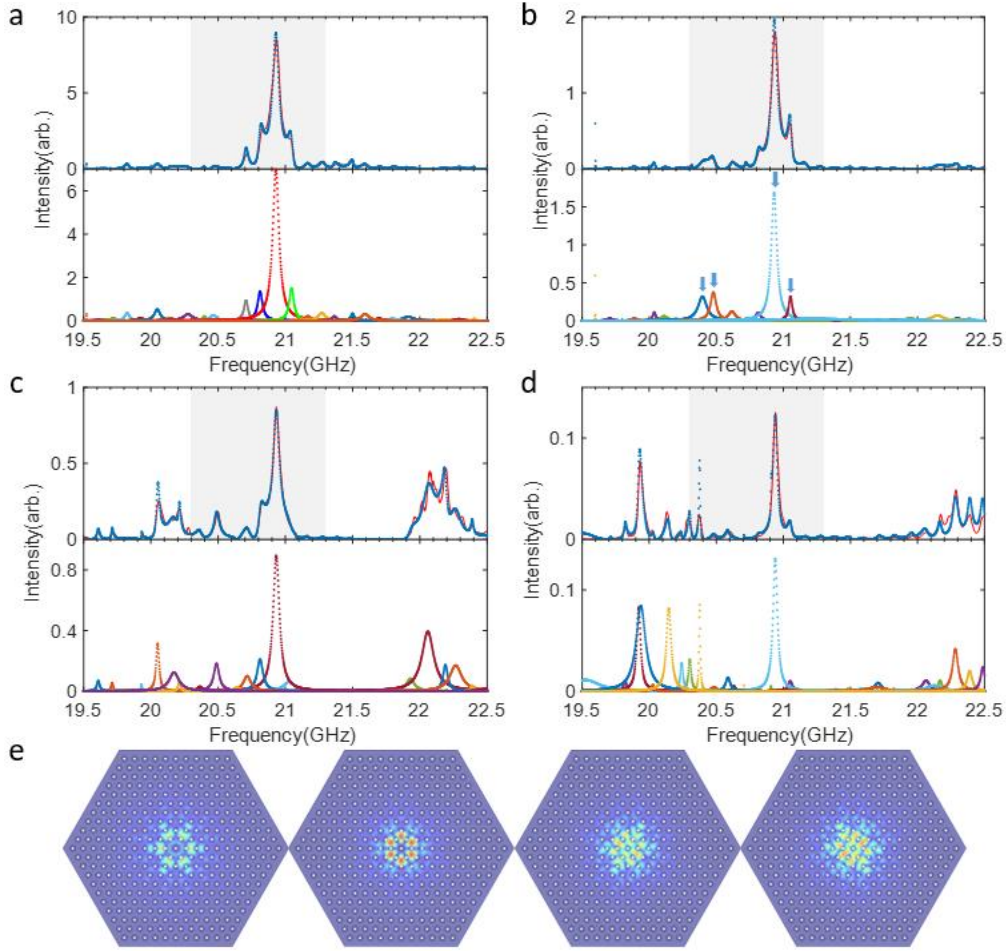

Fig.2. Measurement of cavity modes. (a~d) Upper panel: The red curve is the measured spectrum, the blue dotted curve is the fit result based on harmonic inversion method. The band gap is indicated by the shaded region. Lower panel: The contribution of each of the quasi-normal modes. The distances between the cavity and the boundary line is 1/3/5/7 layers in a-d, respectively. (e) Numerical calculation of the eigenmodes. From left to right, the modes correspond to the four modes indicated by arrows in the lower panel of (b).

Table 1. Complex central frequency $\omega_n - i\Gamma_{n\text{tot}}$ of modes extracted from measurement (GHz)

| Numeric eigenmode | Measurement | | | |
|---|---|---|---|---|
| | a | b | c | d |
| 20.376 | 20.397-0.014i | 20.397-0.039i | 20.364-0.0237i | 20.376-0.0054i |
| 20.502 | 20.464-0.0268i | 20.478-0.0251i | 20.488-0.0247i | 20.481-0.0253i |
| 20.571 | 20.707-0.0187i | 20.616-0.0305i | 20.716-0.0336i | 20.587-0.0199i |
| 20.572 | 20.812-0.0217i | 20.812-0.0243i | 20.811-0.0262i | |
| 21.043 | 20.922-0.028i | 20.931-0.0248i | 20.932-0.0284i | 20.939-0.0198i |
| 21.046 | 21.046-0.0207i | 21.053-0.0164i | 21.027-0.0488i | 21.052-0.0184i |

The 1st/2nd/5th/6th eigenmodes can be observed in measurements. Intensity patterns of these eigenmodes are shown in Fig. 2e. The 3rd/4th mode are not detected in the experiment. Some measured modes, such as the mode at 20.707GHz in column a, only appear once in the measurement. When the cavity is moved to another position, this mode disappear. Such kind of modes related to the local imperfections inside the cavity.

Spectra of the intensity transmission coefficient and the phase shift are shown in Figs. 3a,b. An abrupt phase change of -0.6$\pi$ rad is observed at 21.06GHz, which is a signature of a Fano resonance. The intensity is also near zero at this point. The mode analysis described above shows that the transmission spectrum is the result of the interference between the edge mode and the cavity mode at 20.922-0.028i GHz. The neighboring small peaks are due to other nearby modes.

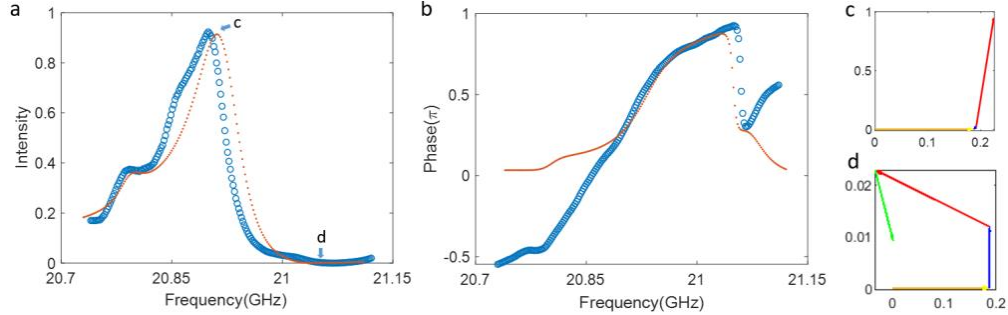

Fig. 3. Transmission spectra. (a) Transmission coefficient with (b) phase variation at the edge on the output. The blue and red curves indicate the measurement and analytical results, respectively. (c,d) Complex plane representation of individual mode contributions at the frequency indicated by arrows c and d in (a). The orange vector indicates the edge channel, while the blue/red/green vectors correspond to the modes represented by the blue/red/green dashed lines at around 21GHz in the lower panel of Fig. 2a.

We first consider the case of a single localized mode and simplify Eq. (2) to

$$\tau = t + \frac{2\sqrt{\Gamma_1 \Gamma_2}}{i(\omega_0 - \omega) + \Gamma_{tot}} e^{i\varphi}. \quad (4)$$

The subscript n is omitted. Since we focus on a narrow frequency range, $t$ can be taken to be a constant real number.

Scattering matrix is not unitary in a dissipative system, we need another constraint to determine $\varphi$. Yoon et al. point out the quasi-reversibility [19] of the scattering matrix as

follows: in a dissipative system, $|s_-\rangle = S_{loss}|s_+\rangle$, from Eq. (1) we obtain

$$S_{loss} = C + \frac{KK^T}{i(\omega_0 - \omega) + \Gamma_1 + \Gamma_2 + \Gamma_{int}}.$$ In the time reversal case, $|s_+\rangle^*$ will be scattered to $|s_-\rangle^*$ while the internal loss is replaced by gain, so that $|s_+\rangle^* = S_{gain}|s_-\rangle^*$, by replacing $\Gamma_{int}$ by $-\Gamma_{int}$, we get $$S_{gain} = C + \frac{KK^T}{i(\omega_0 - \omega) + \Gamma_1 + \Gamma_2 - \Gamma_{int}}.$$ Because $S_{loss}^* = S_{gain}^{-1}$, this relation uniquely determines the phase difference $\varphi$ between the edge mode term $t$ and resonance term in Eq. (4), and yields [19]

$$\cos\varphi = -\frac{(\Gamma_1 + \Gamma_2)t}{2\sqrt{\Gamma_1\Gamma_2}} \quad (5)$$

The interference profile between a continuum state and a single discrete mode can be expressed as $|\tau|^2 = |t|^2 \frac{(q+\delta)^2}{\delta^2+1}$ [10], with the normalized frequency, $\delta = (\omega - \omega_0)/\Gamma$, $\omega_0$ is the central frequency and $\Gamma$ is the half linewidth of the resonant state, and $q$ is shape parameter determined by the coupling between resonant and non-resonant modes. $q$ is obtained from the frequency difference between the central frequency of the discrete mode and the zero intensity point in this spectrum. The value of $t$ is estimated from the peak signal $T_{max} = |t|^2(q^2+1)$. The final analytical fit is shown in Figs. 3a,b. Without loss, the phase change would be $-\pi$ and the transmission would vanish. In the measurement, the phase change is affected by the internal dissipation and by presence of other modes. We fit the data using a model of three cavity modes at 20.812/20.922/21.046 GHz. The contribution of each mode at the peak and valley of the signal is shown in Figs. 3c,d. The edge mode and resonance mode interfere destructively at the minimum in intensity. The analytical results ignore the nearby modes, so that the phase at two ends does not align with the measurement.

We consider the quasi-normal mode at 20.922 GHz. It should be noticed that the parameters obtained in the fit are not uniquely determined, however, $\Gamma_1$ cannot be zero. Comparing Eq. (4) with the standard Fano expression, we obtain $q^2 + 1 = \frac{4\Gamma_1\Gamma_2}{\Gamma_{tot}^2 t^2}$ [19]. Thus

$$\frac{\Gamma_1}{\Gamma_{tot}} > \frac{1 - \sqrt{1-(q^2+1)t^2}}{2} \sim 0.3.$$ We conclude that at least 30 percent of the linewidth of the cavity mode is due to coupling to the backward channel $s_{1-}$. Since the forward channel and backward channels only support the spin-up-polarized and spin-down-polarized mode, respectively, the large value of $\frac{\Gamma_1}{\Gamma_{tot}}$ indicates that cavity is coupled to both the forward and backward edge channel and pseudo-spin polarization is not conserved for light trapped inside

the cavity. Based on the Hamiltonian approximation mentioned previously, the disorder introduced here is proportional to $s_z$, so that this kind of disorder should not mix the spin up and spin down states. Additionally, the spin flipping should be inhibited as long as the disorder has TR symmetry. Our measurement shows this Hamiltonian approximation does not works well for the whole bandgap in the presence of a large resonant cavity. Spin coupling is not negligible.

The delay time in transmission at a given frequency, which is the delay of a pulse in the limit of vanishing bandwidth, is equal to the derivative of the phase of the transmitted field with angular frequency $d\phi/d\omega$ [20]. The negative phase derivative at the zero-intensity point indicates the pulse is reshaped at this frequency.

## 4. Generalization to system broken TR symmetry

The analysis of the transmission spectrum shows that reflection is present in this TI system. However, transmission can be robust in a nonreciprocal system, for instance, chiral edge states in gyromagnetic photonic crystals [21–23]. We generalize this argument to a case in which only the forward channel can be supported.

In the case of a single discrete mode, $D$ becomes the outgoing coupling coefficient $d$ while $K$ becomes the incoming coupling coefficient $k$. Mann et al. [24] demonstrate that $td^* = -k$ for a nonreciprocal system. Equation (2) can be rewritten to give [25]

$$\tau = t + \frac{dk}{i(\omega_0 - \omega) + \Gamma_{tot}} = t(1 - \frac{2\Gamma_{rad}}{i(\omega_0 - \omega) + \Gamma_{tot}}) = t(1 - \frac{2\eta}{1 - i\delta}), \quad (6)$$

where $\Gamma_{rad} = \frac{dd^*}{2}$ is the linewidth due to outgoing coupling, $\eta = \frac{\Gamma_{rad}}{\Gamma_{tot}}$ and $\delta = (\omega - \omega_0)/\Gamma_{tot}$.

The complex representation of $\tau$ is a circle centered at $((1-\eta)t, 0)$ with radius $\eta t$, as shown in Fig. 4a. As shown in Figs. 4b,c, when $\eta = 0.5$, the circle crosses the origin, and there is an abrupt phase change of $-\pi$. For $\eta > 0.5$, the phase increases by $2\pi$ through the mode. In contrast, the phase derivative will be negative for $\eta < 0.5$.

We further check this in the tight-binding model, the effect of a cavity was emulated in the lossless system ($\eta = 1$) following the Haldane model [26]. Numerical results are obtained using an open-source package Kwant [27]. The transmission is unity since there is no backward propagating channel. We therefore focus on the phase change. The phase variation of the transmitted field is shown in Fig. 4d. The phase increases by $2\pi$ twice, indicating that the cavity supports two modes within this frequency range. The speckle patterns at two central frequencies of these two modes are plotted in the inset. In this case, the phase derivative equals the intensity integral of the system divided by $2\pi$.

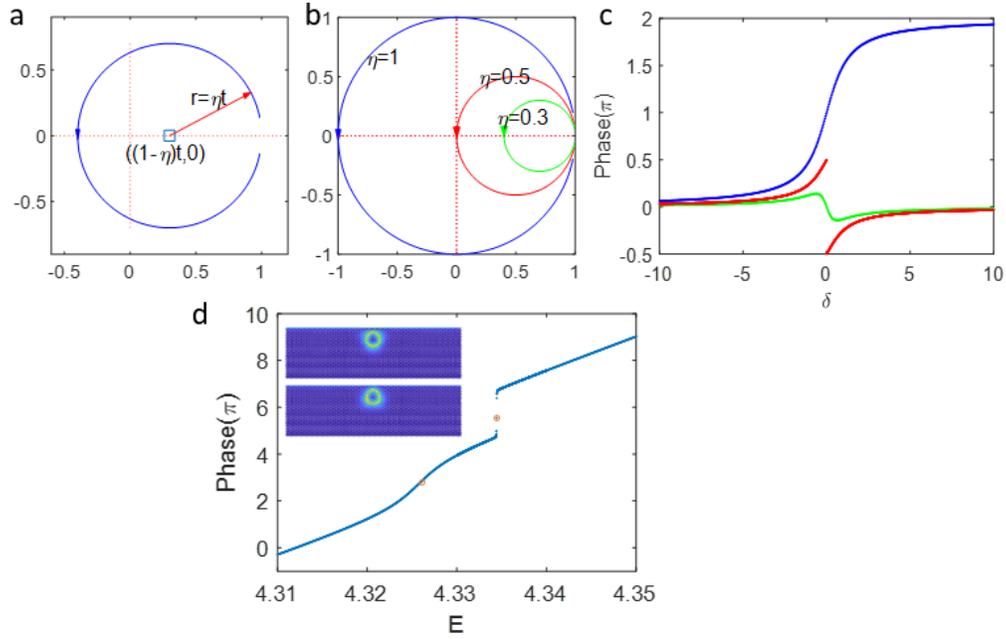

Fig. 4. Transmission through a nonreciprocal system. (a) General complex representation of transmission and (b) complex representation for three cases with $\eta =1/0.5/0.3$, with phase variation as shown in (c). The arrow represents the direction of increasing $\delta$. (d) Tight binding simulation following the Haldane model $H = \sum_i c_i^\dagger \eta_i c_i + \sum_{\langle ij \rangle} c_i^\dagger t c_j + \sum_{\langle\langle ij \rangle\rangle} c_i^\dagger i\lambda_{SO} v_{ij} c_j$, for onsite energy $\eta_i = 4$, nearest-neighbor hopping $t = -1$, and spin-obit coupling strength $\lambda_{SO} = 0.1$. Inside the cavity, $\lambda_{SO} = -0.1$. $v_{ij} = \pm 1$. The blue line is the phase change of transmitted field. Two discrete modes are excited. The inset shows the intensity distribution at two energies indicated by red dots.

## 5. Conclusion

We have observed a Fano resonance between a continuum edge state and an extended defect in a time-reversal invariant TI structure. Our results contrast with previous work, in which the edge state resisted spin flipping in the presence of point defects [6,12]. The time delay near resonance increases with increasing energy inside the system. In the present work, we find backscattering induced by an extended cavity, which accounts for 30% of the linewidth of the cavity mode. Thus it is not possible to increase the transport time while maintaining perfect transmission in a bianisotropic structure with TR symmetry by introducing a cavity. However, this cavity design can be utilized in systems without reciprocity. Based on the coupled-mode theory, as long as decay rate due to internal dissipation is smaller than the rate of coupling, the phase change is always positive. Transmitted pulses can be modified and the time delay lengthened by introducing a cavity structure in nonreciprocal systems. This work points the way to a broader class of disordered TI system in which the edge mode is coupled to extended defects along the length of the edge. The defects may be extended cavities arranged either periodically or randomly, or the systems with random disorder in which spatially localized modes are introduced into the band gap as a result of Anderson localization.

**Acknowledgments**

The work of Y.K. and A.Z.G. is supported by the National Science Foundation (NSF/DMR/-BSF: 1609218). We thank Xiang Ni and Alexander B. Khanikaev for designing the bianisotropic structure.